\documentclass[letterpaper,12pt]{article}
\usepackage{tabularx} % extra features for tabular environment
\usepackage{amsmath}  % improve math presentation
\usepackage{graphicx} % takes care of graphic including machinery
\usepackage[margin=1in,letterpaper]{geometry} % decreases margins
\usepackage{cite} % takes care of citations
\usepackage[final]{hyperref} % adds hyper links inside the generated pdf file
\hypersetup{
	colorlinks=true,       % false: boxed links; true: colored links
	linkcolor=blue,        % color of internal links
	citecolor=blue,        % color of links to bibliography
	filecolor=magenta,     % color of file links
	urlcolor=blue         
}
\usepackage{mathrsfs}%花体字母
\usepackage{float}

\usepackage{revsymb}

\usepackage{amssymb}

\usepackage{exscale}

\usepackage{relsize}

\usepackage{gensymb}

\begin{document}

\title{Theorems on Transverse-Longitudinal Coupling-Based Bunch Compression and Harmonic Generation Schemes}
%\author{Xiujie Deng,  Zizheng Li, Yao Zhang, Zhilong Pan,\\
%Alex Chao, Chuanxiang Tang, Tsinghua University, Beijing, China.}
\author{Xiujie Deng\thanks{dengxiujie@mail.tsinghua.edu.cn}, Tsinghua University, Beijing, China}
\date{\today}
\maketitle

\begin{abstract}
In particle accelerators, transverse-longitudinal coupling (TLC) dynamics can be invoked for efficient bunch compression or high harmonic generation  when one of the transverse eigenemittance is small. In this sense, complete or partial transverse-to-longitudinal emittance exchange in optical wavelength range is being actively studied, for example in free-electron lasers~\cite{Cornacchia2002,Emma2006,Xiang2010,Jiang2011,Xiang2011,Deng2013,Feng2014,Feng2017,Wang2019,Wang2020,Lu2022}. Another example is the recent work on generalized longitudinal strong focusing steady-state microbunching~\cite{Li2023}, where TLC is exploited to take advantage of the ultrasmall vertical emittance in a planar electron storage ring to lower the modulation laser power for ultrashort microbunch generation on a turn-by-turn basis. For this kind of schemes, we have proved three theorems in Ref.~\cite{Deng2023,Deng2021}, invoking 4D phase space dynamics, with their implications discussed. Here we generalize the analysis to 6D phase space dynamics. Various TLC-based beam manipulation scenarios, as listed in the references, are dictated by these theorems. 
%The work presented is useful in practical applications of these schemes. 
\end{abstract}

%\section{Bunching factor of ADM}

%We assume that $\lambda_{\text{L}}=1064$ nm, and the desired radiation wavelength $\lambda_{r}=50\sim150$~nm.  Both modulator and radiator are assumed to be planar undulators.

If the initial bunch is longer than the modulation radiofrequency ( RF) or laser  wavelength, then compression of bunch or microbunch can just be viewed as a harmonic generation scheme. Therefore, in this paper, we will treat bunch compression and harmonic generation as the same thing. 

\begin{figure}[H]
	\centering 
	\includegraphics[width=1\textwidth]{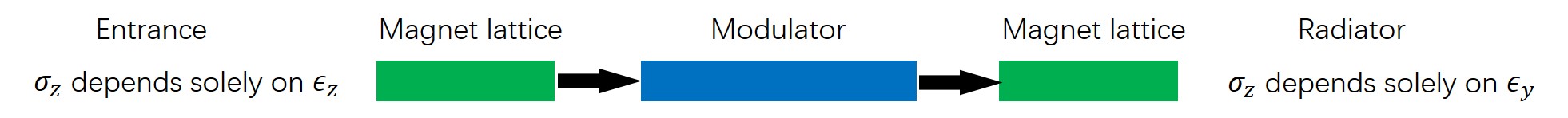}
	\caption{
		\label{fig:Chap3-TLC} % spaces are big no-no withing labels
		A schematic layout of applying TLC dynamics for bunch compression.}
\end{figure}

\section{Problem Definition}
Let us first define the problem we are trying to solve. Particle state vector ${\bf X}=\left(x,x',y,y',z,\delta\right)^{T}$ is used, with the superscript $^{T}$ meaning the transpose of a vector or matrix.  We assume $\epsilon_{y}$ is the small  eigenemittance we want to exploit. 
The case of using $\epsilon_{x}$ is similar.  The schematic layout of a TLC-based bunch compression section is shown in Fig.~\ref{fig:Chap3-TLC}. Suppose the beam at the entrance of the bunch compression section is $x$-$y$-$z$ decoupled, with its second moments matrix given by
\begin{equation}
\Sigma_{i}=\left(\begin{matrix}
\epsilon_{x}\beta_{xi}&-\epsilon_{x}\alpha_{xi}&0&0&0&0\\
-\epsilon_{x}\alpha_{xi}&\epsilon_{x}{\gamma_{xi}}&0&0&0&0\\
0&0&\epsilon_{y}\beta_{yi}&-\epsilon_{y}\alpha_{yi}&0&0\\
0&0&-\epsilon_{y}\alpha_{yi}&\epsilon_{y}{\gamma_{yi}}&0&0\\
0&0&0&0&\epsilon_{z}\beta_{zi}&-\epsilon_{z}\alpha_{zi}\\
0&0&0&0&-\epsilon_{z}\alpha_{zi}&\epsilon_{z}{\gamma_{zi}}\\
\end{matrix}\right),
\end{equation}
where $\alpha$, $\beta$ and $\gamma$ are the Courant-Snyder functions, the subscript $_{i}$ means initial, and $\epsilon_{x}$, $\epsilon_{y}$ and $\epsilon_{z}$ are the eigenemittances of the beam corresponding to the horizontal, vertical and longitudinal mode, respectively. Note that eigenemittances are beam invariants with respect to linear symplectic transport. For the application of TLC for bunch compression, it means that the final bunch length at the exit or radiator $\sigma_{z}(\text{Rad})$ depends only on the vertical emittance $\epsilon_{y}$ and not on the horizontal one $\epsilon_{x}$ and longitudinal one $\epsilon_{z}$.

We divide such a bunch compression section into three parts, with their symplectic transfer matrices given by
\begin{equation}
\begin{aligned}
&{\bf M}_{1}=\left(\begin{matrix}
r_{11}&r_{12}&r_{13}&r_{14}&0&r_{16}\\
r_{21}&r_{22}&r_{23}&r_{24}&0&r_{26}\\
r_{31}&r_{32}&r_{33}&r_{34}&0&r_{36}\\
r_{41}&r_{42}&r_{43}&r_{44}&0&r_{46}\\
r_{51}&r_{52}&r_{53}&r_{54}&1&r_{56}\\
0&0&0&0&0&1\\
\end{matrix}\right),\\ 
&{\bf M}_{2}=\text{modulation kick map},\\ 
&{\bf M}_{3}=\left(\begin{matrix}
R_{11}&R_{12}&R_{13}&R_{14}&0&R_{16}\\
R_{21}&R_{22}&R_{23}&R_{24}&0&R_{26}\\
R_{31}&R_{32}&R_{33}&R_{34}&0&R_{36}\\
R_{41}&R_{42}&R_{43}&R_{44}&0&R_{46}\\
R_{51}&R_{52}&R_{53}&R_{54}&1&R_{56}\\
0&0&0&0&0&1\\
\end{matrix}\right),\\
\end{aligned}
\end{equation}	
with ${\bf M}_{1}$ representing ``from entrance to modulator", ${\bf M}_{2}$ representing ``modulation kick" and ${\bf M}_{3}$ representing ``modulator to radiator". 
Note that ${\bf M}_{1}$ and ${\bf M}_{3}$ are in their general thick-lens form, and does not need to be $x$-$y$ decoupled. The transfer matrix from the entrance to the radiator is then
\begin{equation}
{\bf T}={\bf M}_{3}{\bf M}_{2}{\bf M}_{1}.
\end{equation}	
From the problem definition, for $\sigma_{z}(\text{Rad})$ to be independent of $\epsilon_{x}$ and $\epsilon_{z}$, 
we need
\begin{equation}\label{eq:bunchcompressioncondition2}
\begin{aligned}
T_{51}&=0,\
T_{52}=0,\
T_{55}=0,\
%T_{56}=&d \left(R_{43} D-R_{33} D'\right)+d' \left(R_{44} D-R_{34} D'\right)+r_{56}(h R_{56}+1)+R_{56}=0.
T_{56}= 0.
\end{aligned}
\end{equation}

\section{Theorems}

Given the above problem definition, we have three theorems which dictate the relation between the modulator kick strength with the optical functions at the modulator and radiator, respectively.\\
{\bf Theorem one:} If 
\begin{equation}\label{eq:M21}
{\bf M}_{2}=\left(\begin{matrix}
1&0&0&0&0&0\\
0&1&0&0&0&0\\
0&0&1&0&0&0\\
0&0&0&1&0&0\\
0&0&0&0&1&0\\
0&0&0&0&h&1\\
\end{matrix}\right),
\end{equation}
which corresponds to the case of a normal RF or a TEM00 mode laser modulator,
then 
\begin{equation}\label{eq:theorem1}
h^2(\text{Mod})\mathcal{H}_{y}(\text{Mod})\mathcal{H}_{y}(\text{Rad})\geq1.
\end{equation}
{\bf Theorem two:} If 
\begin{equation}\label{eq:M22}
{\bf M}_{2}=\left(\begin{matrix}
1&0&0&0&0&0\\
0&1&0&0&0&0\\
0&0&1&0&0&0\\
0&0&0&1&t&0\\
0&0&0&0&1&0\\
0&0&t&0&0&1\\
\end{matrix}\right),
\end{equation}
which corresponds to the case of a transverse deflecting  (in $y$-dimension) RF or a TEM01 mode laser modulator or other schemes for angular modulation,
then 
\begin{equation}\label{eq:theorem2}
t^2(\text{Mod})\beta_{y}(\text{Mod})\mathcal{H}_{y}(\text{Rad})\geq1.
\end{equation}
{\bf Theorem three:} If 
\begin{equation}\label{eq:M23}
{\bf M}_{2}=\left(\begin{matrix}
1&0&0&0&0&0\\
0&1&0&0&0&0\\
0&0&1&0&k&0\\
0&0&0&1&0&0\\
0&0&0&0&1&0\\
0&0&0&-k&0&1\\
\end{matrix}\right),
\end{equation}
whose physical correspondence is not as straightforward as the previous two cases, then 
\begin{equation}\label{eq:theorem3}
k^2(\text{Mod})\gamma_{y}(\text{Mod})\mathcal{H}_{y}(\text{Rad})\geq1.
\end{equation}

\section{Proof}\label{sec:proof}
Here we present the details for the proof of Theorem one. The proof of the other two is just similar. From the problem definition, for $\sigma_{z}(\text{Rad})$ to be independent of $\epsilon_{x}$ and $\epsilon_{z}$, 
we need
\begin{equation}\label{eq:TLCCondition2}
\begin{aligned}
T_{51}&=r_{11} R_{51}+r_{21} R_{52}+r_{31} R_{53}+r_{41} R_{54}+r_{51} \left(h R_{56}+1\right)=0,\\
T_{52}&=r_{12} R_{51}+r_{22} R_{52}+r_{32} R_{53}+r_{42} R_{54}+r_{52} \left(h R_{56}+1\right)=0,\\
T_{55}&= h R_{56}+1=0,\\
%T_{56}=&d \left(R_{43} D-R_{33} D'\right)+d' \left(R_{44} D-R_{34} D'\right)+r_{56}(h R_{56}+1)+R_{56}=0.
T_{56}&= r_{16} R_{51}+r_{26} R_{52}+r_{36} R_{53}+r_{46} R_{54}+r_{56} \left(h R_{56}+1\right)+R_{56}=0.
\end{aligned}
\end{equation}
Under the above conditions, we have
\begin{equation}\label{eq:TransportMatrix12}
\begin{aligned}
&{\bf T}=\left(
\begin{matrix}
{\bf A}&{\bf B}&{\bf C}\\
{\bf D}&{\bf E}&{\bf F}\\
{\bf G}&{\bf H}&{\bf I}
\end{matrix}
\right),
\end{aligned}
\end{equation}
with ${\bf A}\sim{\bf I}$ being $2\times2$ submatrices of ${\bf T}$ where
\begin{equation}\label{eq:TransportMatrixSub2}
\begin{aligned}
&{\bf G}=\left(\begin{matrix}
0&0\\
r_{51}h &r_{52}h\\
\end{matrix}\right),\\
&{\bf H}=\left(\begin{matrix}
r_{13} R_{51}+r_{23} R_{52}+r_{33} R_{53}+r_{43} R_{54}&r_{14} R_{51}+r_{24} R_{52}+r_{34} R_{53}+r_{44} R_{54}\\
r_{53}h &r_{54}h\\
\end{matrix}\right),\\
&{\bf I}=\left(\begin{matrix}
0&0\\
h&r_{56}h +1\\
\end{matrix}\right).
\end{aligned}
\end{equation}
The bunch length squared at the modulator and the radiator are
\begin{equation}\label{eq:BL2}
\begin{aligned}
\sigma_{z}^{2}(\text{Mod})
%=&\epsilon_{z}(\beta_{z}-2\alpha_{z}r_{56}+\gamma_{z}r_{56}^2)\\
%&+\epsilon_{y}\left(\beta_{y}r_{53}^2-2\alpha_{y}r_{53}r_{54}+\gamma_{x}r_{54}^2\right)\\
&=\epsilon_{x}\frac{\left(\beta_{xi}r_{51}-\alpha_{xi}r_{52}\right)^2+r_{52}^2}{\beta_{xi}}+\epsilon_{y}\frac{\left(\beta_{yi}r_{53}-\alpha_{yi}r_{54}\right)^2+r_{54}^2}{\beta_{yi}}+\epsilon_{z}\left(\beta_{zi}-2\alpha_{zi}r_{56}+\gamma_{zi}r_{56}^2\right)\\
&=\epsilon_{x}\mathcal{H}_{x}(\text{Mod})+\epsilon_{y}\mathcal{H}_{y}(\text{Mod})+\epsilon_{z}\beta_{z}(\text{Mod}),\\
%&+\epsilon_{y}\left(\beta_{y}\left[r_{43}d-r_{33}d'\right]^2-2\alpha_{y}\left[r_{43}d-r_{33}d'\right]\left[-r_{34}d'+r_{44}d\right]+\gamma_{x}\left[-r_{34}d'+r_{44}d\right]^2\right)\\
\sigma_{z}^{2}(\text{Rad})
%=&\epsilon_{y}\left(\beta_{y}T_{53}^2-2\alpha_{y}T_{53}T_{54}+\gamma_{x}T_{54}^2\right)\\
&=\epsilon_{y}\frac{\left(\beta_{yi}T_{53}-\alpha_{yi}T_{54}\right)^2+T_{54}^2}{\beta_{yi}}
=\epsilon_{y}\mathcal{H}_{y}(\text{Rad}).
%&=\epsilon_{y}(\beta_{y}\left[D \left(r_{33} R_{43}+r_{43} R_{44}\right)-\left(r_{33} R_{33}+r_{43} R_{34}\right) D'\right]^2\\
%&-2\alpha_{y}\left[D \left(r_{33} R_{43}+r_{43} R_{44}\right)-\left(r_{33} R_{33}+r_{43} R_{34}\right) D'\right]\left[D \left(r_{34} R_{43}+r_{44} R_{44}\right)-\left(r_{34} R_{33}+r_{44} R_{34}\right) D'\right]\\
%&+\gamma_{x}\left[D \left(r_{34} R_{43}+r_{44} R_{44}\right)-\left(r_{34} R_{33}+r_{44} R_{34}\right) D'\right]^2)\\
\end{aligned}
\end{equation}
%from which we can see that the generalized $\mathcal{H}_{y}$ function is a parameter quantifying the bunch length contribution from the transverse emittance.
%, at the modulator and radiator are defined as
%\begin{equation}
%\begin{aligned}
%\mathcal{H}_{y}(\text{Mod})&=\frac{\left(\beta_{y}r_{53}-\alpha_{y}r_{54}\right)^2+r_{54}^2}{\beta_{y}},\\ \mathcal{H}_{y}(\text{Rad})&=\frac{\left(\beta_{y}T_{53}-\alpha_{y}T_{54}\right)^2+T_{54}^2}{\beta_{y}}.
%\end{aligned}
%\end{equation}
%The theorem we want to prove is 
%$
%h^2(\text{Mod})\mathcal{H}_{y}(\text{Mod})\mathcal{H}_{y}(\text{Rad})\geq1.
%$
%If we ignore the contribution of $\epsilon_{z}$ on $\sigma_{z}(\text{Mod})$, then the theorem can also be written as
%$
%|h(\text{Mod})|\geq\frac{\epsilon_{y}}{\sigma_{z}(\text{Mod})\sigma_{z}(\text{Rad})}.
%$
According to Cauchy-Schwarz inequality, we have
%\begin{widetext}
\begin{equation}\label{eq:proof}
\begin{aligned}
h^2(\text{Mod})\mathcal{H}_{y}(\text{Mod})\mathcal{H}_{y}(\text{Rad})&=h^2\frac{\left[\left(\beta_{yi}r_{53}-\alpha_{yi}r_{54}\right)^2+r_{54}^2\right]}{\beta_{yi}}\frac{\left[\left(\beta_{yi}T_{53}-\alpha_{yi}T_{54}\right)^2+T_{54}^2\right]}{\beta_{yi}}\\
&\geq
\frac{h^2}{\beta_{yi}^{2}}\left[-\left(\beta_{yi}r_{53}-\alpha_{yi}r_{54}\right)T_{54}+r_{54}\left(\beta_{yi}T_{53}-\alpha_{yi}T_{54}\right)\right]^{2}\\
&=\left(T_{53}r_{54}h-T_{54}r_{53}h\right)^{2}=\left(T_{53}T_{64}-T_{54}T_{63}\right)^{2}=|\text{det}({\bf H})|^{2}.
\end{aligned}
\end{equation}
The equality holds when
$
\frac{-\left(\beta_{yi}r_{53}-\alpha_{yi}r_{54}\right)}{T_{54}}=\frac{r_{54}}{\left(\beta_{yi}T_{53}-\alpha_{yi}T_{54}\right)}.
$
The symplecticity of ${\bf T}$ requires that ${\bf T}{\bf S}{\bf T}^{T}={\bf S}$, where ${\bf S}=\left(
\begin{matrix}
{\bf J}&0&0\\
0&{\bf J}&0\\
0&0&{\bf J}
\end{matrix}
\right)$ and $
{\bf J}=\left(
\begin{matrix}
0&1\\
-1&0
\end{matrix}
\right),$ so we have
\begin{equation}
\left(
\begin{matrix}
{\bf A}{\bf J}{\bf A}^{T}+{\bf B}{\bf J}{\bf B}^{T}+{\bf C}{\bf J}{\bf C}^{T}&{\bf A}{\bf J}{\bf D}^{T}+{\bf B}{\bf J}{\bf E}^{T}+{\bf C}{\bf J}{\bf F}^{T}&{\bf A}{\bf J}{\bf G}^{T}+{\bf B}{\bf J}{\bf H}^{T}+{\bf C}{\bf J}{\bf I}^{T}\\
{\bf D}{\bf J}{\bf A}^{T}+{\bf E}{\bf J}{\bf B}^{T}+{\bf F}{\bf J}{\bf C}^{T}&{\bf D}{\bf J}{\bf D}^{T}+{\bf E}{\bf J}{\bf E}^{T}+{\bf F}{\bf J}{\bf F}^{T}&{\bf D}{\bf J}{\bf G}^{T}+{\bf E}{\bf J}{\bf H}^{T}+{\bf F}{\bf J}{\bf I}^{T}\\
{\bf G}{\bf J}{\bf A}^{T}+{\bf H}{\bf J}{\bf B}^{T}+{\bf I}{\bf J}{\bf C}^{T}&{\bf G}{\bf J}{\bf D}^{T}+{\bf H}{\bf J}{\bf E}^{T}+{\bf I}{\bf J}{\bf F}^{T}&{\bf G}{\bf J}{\bf G}^{T}+{\bf H}{\bf J}{\bf H}^{T}+{\bf I}{\bf J}{\bf I}^{T}\\
\end{matrix}
\right)
=
{\bf S}.
\end{equation}
%\end{widetext}
According to Eq.~(\ref{eq:TransportMatrixSub2}), we have ${\bf G}{\bf J}{\bf G}^{T}=\left(
\begin{matrix}
0&0\\
0&0
\end{matrix}
\right)$, ${\bf I}{\bf J}{\bf I}^{T}=\left(
\begin{matrix}
0&0\\
0&0
\end{matrix}
\right)$. Therefore,
\begin{equation}
{\bf H}{\bf J}{\bf H}^{T}={\bf J},
\end{equation}
which means ${\bf H}$ is also a symplectic matrix. So we have
$\text{det}({\bf H})=1.
$
The theorem is thus proven.

\section{Dragt's Minimum Emittance Theorem}

Theorem one in Eq.~(\ref{eq:theorem1}) can also be expressed as
\begin{equation}
|h(\text{Mod})|\geq\frac{\epsilon_{y}}{\sqrt{\epsilon_{y}\mathcal{H}_{y}(\text{Mod})}\sqrt{\epsilon_{y}\mathcal{H}_{y}(\text{Rad})}}=\frac{\epsilon_{y}}{\sigma_{zy}(\text{Mod})\sigma_{z}(\text{Rad})}.
\end{equation}
Note that in the above formula,  $\sigma_{zy}(\text{Mod})$ means the bunch length at the modulator contributed from the vertical emittance $\epsilon_{y}$. So given a fixed $\epsilon_{y}$ and desired $\sigma_{z}(\text{Rad})$, a smaller $h(\text{Mod})$, i.e., a smaller RF acceleration gradient or modulation laser power ($P_{\text{laser}}\propto|h(\text{Mod})|^2$), means a larger $\mathcal{H}_{y}(\text{Mod})$, thus a longer $\sigma_{zy}(\text{Mod})$, is needed.   As $|h(\text{Mod})|\sigma_{z}(\text{Mod})$ quantifies the energy spread introduced by the modulation kick, we thus also have 
\begin{equation}\label{eq:ES}
\sigma_{z}(\text{Rad})\sigma_{\delta}(\text{Rad})\geq\epsilon_{y}.
\end{equation}
Similarly for Theorem two and three, we have
\begin{equation}
\begin{aligned}
|t(\text{Mod})|&\geq\frac{\epsilon_{y}}{\sigma_{y\beta}(\text{Mod})\sigma_{z}(\text{Rad})},
\end{aligned}
\end{equation}
and 
\begin{equation}
\begin{aligned}
|k(\text{Mod})|&\geq\frac{\epsilon_{y}}{\sigma_{y'\beta}(\text{Mod})\sigma_{z}(\text{Rad})},
\end{aligned}
\end{equation}
respectively, and also Eq.~(\ref{eq:ES}). Note that in the above formulas, the vertical beam size or divergence at the modulator contains only the vertical betatron part, i.e., that from the vertical emittance $\epsilon_{y}$.

Equation~(\ref{eq:ES}) is actually a manifestation of the classical uncertainty principle \cite{Dragt2020}, which states that 
\begin{equation}
\begin{aligned}
\Sigma_{11}\Sigma_{22}&\geq\epsilon_{\text{min}}^{2},\\
\Sigma_{33}\Sigma_{44}&\geq\epsilon_{\text{min}}^{2},\\
\Sigma_{55}\Sigma_{66}&\geq\epsilon_{\text{min}}^{2},
\end{aligned}
\end{equation}
in which $\epsilon_{\text{min}}$ is the minimum one among the three eigen emittances $\epsilon_{I,II,III}$. In our bunch compression case, we assume that $\epsilon_{y}$ is the smaller one compared to $\epsilon_{z}$. Actually there is a stronger inequality compared to the classical uncertainty principle, i.e., the minimum emittance theorem \cite{Dragt2020}, which states that the projected emittance cannot be smaller than the minimum one among the three eigen emittances,
\begin{equation}
\begin{aligned}
\epsilon_{x,\text{pro}}^{2}=\Sigma_{11}\Sigma_{22}-\Sigma_{12}^{2}&\geq\epsilon_{\text{min}}^{2},\\
\epsilon_{y,\text{pro}}^{2}=\Sigma_{33}\Sigma_{44}-\Sigma_{34}^{2}&\geq\epsilon_{\text{min}}^{2},\\
\epsilon_{z,\text{pro}}^{2}=\Sigma_{55}\Sigma_{66}-\Sigma_{56}^{2}&\geq\epsilon_{\text{min}}^{2}.
\end{aligned}
\end{equation} 

\section{Theorems Cast in Another Form}
As another way to appreciate the result, here we cast the theorems in a form using the generalized beta functions as introduced in the following Sec.~\ref{sec:beta}. According to definition, we have
\begin{equation}
\mathcal{H}_{y}\equiv\beta_{55}^{II},\ \beta_{y}\equiv\beta_{33}^{II},\ \gamma_{y}\equiv\beta_{44}^{II}.
\end{equation}
\\
{\bf Theorem one:} If ${\bf M}_{2}$ is as shown in Eq.~(\ref{eq:M21}),
then 
\begin{equation}\label{eq:theorem12}
M_{65}^2(\text{Mod})\beta_{55}^{II}(\text{Mod})\beta_{55}^{II}(\text{Rad})\geq1,
\end{equation}
where $M_{65}$ is the $_{65}$ matrix term of ${\bf M}_{2}$, i.e., $h$.\\
{\bf Theorem two:} If ${\bf M}_{2}$ is as shown in Eq.~(\ref{eq:M22}),
then 
\begin{equation}\label{eq:theorem22}
M_{63}^2(\text{Mod})\beta_{33}^{II}(\text{Mod})\beta_{55}^{II}(\text{Rad})\geq1.
\end{equation}
{\bf Theorem three:} If ${\bf M}_{2}$ is as shown in Eq.~(\ref{eq:M23}),
then 
\begin{equation}\label{eq:theorem32}
M_{64}^2(\text{Mod})\beta_{44}^{II}(\text{Mod})\beta_{55}^{II}(\text{Rad})\geq1.
\end{equation}
At the entrance, the generalized Twiss matrix corresponding to eigen mode $I$ is
\begin{equation}
{\bf T}_{I}(\text{Ent})=\left(\begin{matrix}
\beta_{xi}&-\alpha_{xi}&0&0&0&0\\
-\alpha_{xi}&\gamma_{xi}&0&0&0&0\\
0&0&0&0&0&0\\
0&0&0&0&0&0\\
0&0&0&0&0&0\\
0&0&0&0&0&0\\
\end{matrix}\right),
\end{equation}
and similar expressions for ${\bf T}_{II,III}(\text{Ent})$, with $x$ replaced by $y,z$ and the location of the $2\times2$ matrix shifted in the diagonal direction.
Then
\begin{equation}
\beta_{33}^{II}(\text{Mod})=\frac{\left(\beta_{yi}r_{33}-\alpha_{yi}r_{34}\right)^2+r_{34}^2}{\beta_{yi}},
\end{equation}
\begin{equation}
\beta_{44}^{II}(\text{Mod})=\frac{\left(\beta_{yi}r_{43}-\alpha_{yi}r_{44}\right)^2+r_{44}^2}{\beta_{yi}},
\end{equation}
\begin{equation}
\beta_{55}^{II}(\text{Mod})=\frac{\left(\beta_{yi}r_{53}-\alpha_{yi}r_{54}\right)^2+r_{54}^2}{\beta_{yi}},
\end{equation}
\begin{equation}
\beta_{55}^{I}(\text{Rad})=\frac{\left(\beta_{xi}T_{51}-\alpha_{xi}T_{52}\right)^2+T_{52}^2}{\beta_{xi}},
\end{equation}
\begin{equation}
\beta_{55}^{II}(\text{Rad})=\frac{\left(\beta_{yi}T_{53}-\alpha_{yi}T_{54}\right)^2+T_{54}^2}{\beta_{yi}},
\end{equation}
\begin{equation}
\beta_{55}^{III}(\text{Rad})=\frac{\left(\beta_{zi}T_{55}-\alpha_{zi}T_{56}\right)^2+T_{56}^2}{\beta_{zi}}.
\end{equation}
For $\sigma_{z}(\text{Rad})$ to be independent of $\epsilon_{x}$ and $\epsilon_{z}$, we need $\beta_{55}^{I}(\text{Rad})=0$ and $\beta_{55}^{III}(\text{Rad})=0$, which then lead to Eq.~(\ref{eq:bunchcompressioncondition2}). And the following proof procedures are the same as that shown in the above Sec.~\ref{sec:proof}.

%\section{Extended Theorems on TLC-based Transverse Beam Size Shaping}
%
%\begin{figure}[H]
%	\centering 
%	\includegraphics[width=1\textwidth]{TLC2.jpg}
%	%	\includegraphics[width=0.48\columnwidth]{Chap3-BFCorrection2.pdf}
%	\caption{
%		\label{fig:Chap3-TLC2} % spaces are big no-no withing labels
%		A schematic layout of applying TLC dynamics for transverse size shaping.}
%\end{figure}
%
%If we apply TLC for transverse beam size shaping, as defined in Fig.~\ref{fig:Chap3-TLC2}, there are three corresponding theorems.\\
%{\bf Theorem one:} If ${\bf M}_{2}$ is as shown in Eq.~(\ref{eq:M21}),
%then 
%\begin{equation}\label{eq:theorem122}
%M_{65}^2(\text{Mod})\beta_{55}^{III}(\text{Mod})\beta_{33}^{III}(\text{Rad})\geq1.
%\end{equation}
%{\bf Theorem two:} If ${\bf M}_{2}$ is as shown in Eq.~(\ref{eq:M22}),
%then 
%\begin{equation}\label{eq:theorem222}
%M_{63}^2(\text{Mod})\beta_{33}^{III}(\text{Mod})\beta_{33}^{III}(\text{Rad})\geq1.
%\end{equation}
%{\bf Theorem three:} If ${\bf M}_{2}$ is as shown in Eq.~(\ref{eq:M23}),
%then 
%\begin{equation}\label{eq:theorem322}
%M_{64}^2(\text{Mod})\beta_{44}^{III}(\text{Mod})\beta_{33}^{III}(\text{Rad})\geq1.
%\end{equation}

\section{Generalized Beta Functions}\label{sec:beta}

Following Chao's solution by linear matrix (SLIM) formalism~\cite{Chao1979}, we can introduce the definition of the generalized beta functions in a 3D general coupled storage ring lattice as
\begin{equation}
\beta_{ij}^{k}=2\text{Re}\left({\bf E}_{ki}{\bf E}_{kj}^{*}\right),\ k=I,II,III,
\end{equation}
where $^{*}$ means complex conjugate, the sub or superscript $k$ denotes one of the three eigenmodes, Re() means the real component of a complex number or matrix, ${\bf E}_{ki}$ is the $i$-th component of vector  ${\bf E}_{k}$, and ${\bf E}_{k}$ are eigenvectors of the $6\times6$ symplectic one-turn map ${\bf M}$ with eigenvalues $e^{i2\pi\nu_{k}}$, satisfying the following normalization condition
\begin{equation}\label{eq:norm}
{\bf E}_{k}^{\dagger}{\bf S}{\bf E}_{k}=\begin{cases}
&i,\ k=I,II,III,\\
&-i,\ k=-I,-II,-III,
\end{cases}
\end{equation}
and 
${\bf E}_{k}^{\dagger}{\bf S}{\bf E}_{j}=0$ for $k\neq j$, where $^{\dagger}$ means complex conjugate transpose, and ${\bf S}=\left(
\begin{matrix}
{\bf J}&0&0\\
0&{\bf J}&0\\
0&0&{\bf J}
\end{matrix}
\right)$  with $
{\bf J}=\left(
\begin{matrix}
0&1\\
-1&0
\end{matrix}
\right)$.
%For a bounded motion, the eigenvalues lie on a unit circle in complex plane and we have
%\begin{equation}
%\nu_{-k}=-\nu_{k},\ {\bf E}_{-k}={\bf E}_{k}^{*}.
%\end{equation}
%We then have
%\begin{equation}
%\nu_{-k}=-\nu_{k},\ {\bf E}_{-k}={\bf E}_{k}^{*}.
%\end{equation}
%Similar definition has also been adopted in Ref.~\cite{Wolski2006}. 
%Using the generalized beta function, we can write the eigenvector component as
%\begin{equation}
%{\bf E}_{kj}=\sqrt{\frac{\beta_{jj}^{k}}{2}}e^{i\phi_{j}^{k}}.
%\end{equation}
%%Using the generalized beta functions, we can write the normalized eigenvectors as
%%\begin{equation}
%%\begin{aligned}
%%{\bf E}_{I,II,III}&=\frac{1}{\sqrt{2}}\left(\begin{matrix}
%%\sqrt{\beta_{11}^{I,II,III}}e^{i\phi_{1}^{I,II,III}}\\
%%\sqrt{\beta_{22}^{I,II,III}}e^{i\phi_{2}^{I,II,III}}\\
%%\sqrt{\beta_{33}^{I,II,III}}e^{i\phi_{3}^{I,II,III}}\\
%%\sqrt{\beta_{44}^{I,II,III}}e^{i\phi_{4}^{I,II,III}}\\
%%\sqrt{\beta_{55}^{I,II,III}}e^{i\phi_{5}^{I,II,III}}\\
%%\sqrt{\beta_{66}^{I,II,III}}e^{i\phi_{6}^{I,II,III}}\\
%%\end{matrix}\right).
%%\end{aligned}
%%\end{equation}
%And according to definition we have
%\begin{equation}
%\beta_{ij}^{k}=\sqrt{\beta_{ii}^{k}\beta_{jj}^{k}}\cos(\phi_{i}^{k}-\phi_{j}^{k}).
%\end{equation}

Similarly, we introduce the definition of imaginary generalized beta functions as
\begin{equation}
\hat\beta_{ij}^{k}=2\text{Im}\left({\bf E}_{ki}{\bf E}_{kj}^{*}\right),\ k=I,II,III,
\end{equation}
where Im() means the imaginary component of a complex number or matrix. Further we can define the real and imaginary generalized Twiss matrices of a storage ring lattice corresponding to three eigen mode as
\begin{equation}
\left({\bf T}_{k}\right)_{ij}=\beta_{ij}^{k},\ \left(\hat{\bf T}_{k}\right)_{ij}=\hat\beta_{ij}^{k},\ k=I,II,III.
\end{equation}
Due to the symplecticity of the one-turn map, we have
\begin{equation}
\begin{aligned}
%{\bf T}_{-k}&={\bf T}_{k},\ \hat{\bf T}_{-k}=-\hat{\bf T}_{k},\\
{\bf T}_{k}^{T}&={\bf T}_{k},\ \hat{\bf T}_{k}^{T}=-\hat{\bf T}_{k}.
\end{aligned}
\end{equation}
The generalized Twiss matrices at different places are related according to
\begin{equation}
\begin{aligned}
{\bf T}_{k}(s_{2})&={\bf R}(s_{2},s_{1}){\bf T}_{k}(s_{1}){\bf R}^{T}(s_{2},s_{1}),\\ \hat{\bf T}_{k}(s_{2})&={\bf R}(s_{2},s_{1})\hat{\bf T}_{k}(s_{1}){\bf R}^{T}(s_{2},s_{1}),
\end{aligned}
\end{equation}
with ${\bf R}(s_{2},s_{1})$ being the transfer matrix from $s_{1}$ to $s_{2}$.
%The relation between the generalized Twiss matrices and projection matrices is
%\begin{equation}
%\begin{aligned}
%{\bf T}_{k}\equiv2\text{Im}\left({\bf P}_{k}\right){\bf S},\ \hat{\bf T}_{k}\equiv-2\text{Re}\left({\bf P}_{k}\right){\bf S},\ k=I,II,III.
%\end{aligned}
%\end{equation}

The action or generalized Courant-Snyder invariants of a particle are defined as
\begin{equation}
J_{k}\equiv\frac{{\bf X}^{T}{\bf G}_{k}{\bf X}}{2},\ k=I,II,III,
\end{equation}
where 
\begin{equation}
{\bf G}_{k}\equiv{\bf S}^{T}{\bf T}_{k}{\bf S}.
\end{equation}
It is easy to prove that $J_{k}$ are invariants of a particle when it travels around the ring, from the symplectic condition of transfer matrix ${\bf R}^{T}{\bf S}{\bf R}={\bf S}$. The three eigenemittance of a beam containing $N_{p}$ particles are defined according to
\begin{equation}
\epsilon_{k}\equiv\langle J_{k}\rangle=\frac{\sum_{i=1}^{N_{p}}J_{k,i}}{N_{p}},\ k=I,II,III,
\end{equation}
where $J_{k,i}$ means the $k$-th mode invariant of the $i$-th particle.

Assume there is a perturbation ${\bf K}$ to the one-turn map ${\bf M}$, i.e., ${\bf M}_{\text{per}}=({\bf I}+{\bf K}){\bf M}_{\text{unp}}$. From cannonical perturbation theory~\cite{Deng2023per},  the tune shift of the $k$-th eigen mode is then
\begin{equation}
\Delta\nu_{k}=-\frac{1}{4\pi}\text{Tr}\left[\left({\bf T}_{k}+i\hat{\bf T}_{k}\right){\bf S}{\bf K}\right],
\end{equation}
where  Tr() means the trace of a matrix.
This formula can be used to calculate the real and imaginary tune shifts due to symplectic (for example lattice error) and non-symplectic (for example radiation damping) pertubrations.  
%\begin{equation}
%\Delta\nu_{k}=-\frac{i}{2\pi}\text{Tr}({\bf P}_{k}{\bf K})
%=-\frac{1}{4\pi}\text{Tr}\left[\left({\bf T}_{k}+i\hat{\bf T}_{k}\right){\bf S}{\bf K}\right]
%\end{equation}
%\begin{equation}
%{\bf P}_{k}=\frac{\hat{\bf T}_{k}-i{\bf T}_{k}}{2}{\bf S}
%\end{equation}
The pertubation theory can also be applied to calcuate the emittance growth due to diffusion~\cite{Deng2023per}. 
%\begin{equation}
%\Delta\nu_{k}=-\frac{i}{2\pi}\text{Tr}({\bf P}_{k}{\bf K})
%=-\frac{1}{4\pi}\text{Tr}\left[\left({\bf T}_{k}+i\hat{\bf T}_{k}\right){\bf S}{\bf K}\right]
%\end{equation}
%\begin{equation}
%{\bf P}_{k}=\frac{\hat{\bf T}_{k}-i{\bf T}_{k}}{2}{\bf S}
%\end{equation}
With the help of real and imaginary generalized beta functions and Twiss matrices, the diffusion of emittance per turn can be calculated  as
\begin{equation}
\begin{aligned}
\Delta \epsilon_{k}&=-\frac{1}{2}\oint\text{Tr}\left({\bf T}_{k}{\bf S}{\bf N}{\bf S}\right)ds=\frac{1}{2}\oint\text{Tr}\left({\bf G}_{k}{\bf N}\right)ds,\\
%&=-\frac{1}{2}\sum_{i,j}\oint\beta^{k}_{ij}\left({\bf S}{\bf N}{\bf S}\right)_{ij}ds,
\end{aligned}
\end{equation}
and the damping rate of each eigen mode is
\begin{equation}\label{eq:dampC}
\alpha_{k}=-\frac{1}{2}\oint\text{Tr}\left(\hat{\bf T}_{k}{\bf S}{\bf D}\right)ds,
%=\frac{1}{2}\sum_{i,j}\oint\hat{\beta}^{k}_{ij}\left({\bf S}{\bf D}\right)_{ij} ds,
\end{equation}
where ${\bf N}$ and ${\bf D}$ are the diffusion and damping matrix, respectively. Note that the damping rates here are that for the corresponding eigenvectors. The damping rates for particle action or beam emittance is a factor of two larger.  The equilibrium eigenemittance between a balance of diffusion and damping can be calculated as
\begin{equation}
\begin{aligned}
\epsilon_{k}
&=\frac{\Delta\epsilon_{k}}{2\alpha_{k}}
%=\frac{-\frac{1}{2}\oint\text{Tr}\left({\bf T}_{k}{\bf S}{\bf N}{\bf S}\right)ds}{-\oint\text{Tr}\left(\hat{\bf T}_{k}{\bf S}{\bf D}\right)ds}
=\frac{-\frac{1}{2}\sum_{i,j}\oint\beta^{k}_{ij}\left({\bf S}{\bf N}{\bf S}\right)_{ij}ds}{\sum_{i,j}\oint\hat{\beta}^{k}_{ij}\left({\bf S}{\bf D}\right)_{ij} ds},
\end{aligned}
\end{equation} 
After getting the equilibrium eigenemittances, the second moments of beam can be written 
\begin{equation}
\Sigma_{ij}=\sum_{k=I,II,III}\epsilon_{k}\beta_{ij}^{k},
\end{equation}
or in matrix form as
\begin{equation}
{\bf \Sigma}=\sum_{k=I,II,III}\epsilon_{k}{\bf T}_{k}.
\end{equation}
%The above formalsim is general, applies to 3D general coupled lattice, and can be applied for a varity of diffusion and damping mechanisms. 
%This development can be viewed as an extension and generalization of Chao's SLIM formalism~\cite{chao1979evaluation}.
%We can generalize the definition of the chromatic $\mathcal{H}$ and longitudinal beta function in a 3D general coupled lattice as
%\begin{equation}
%\begin{aligned}
%\mathcal{H}_{x}&=\beta_{55}^{I}=2|{\bf E}_{I5}|^2,\\ \mathcal{H}_{y}&=\beta_{55}^{II}=2|{\bf E}_{II5}|^2,\\
%\beta_{z}&=\beta_{55}^{III}=2|{\bf E}_{III5}|^2.
%\end{aligned}
%\end{equation}
%They quantify the contribution of horizontal, vertical and longitudinal emittance to bunch length, respectively.
%\begin{equation}
%\sigma_{z}=\sqrt{\epsilon_{x}\mathcal{H}_{x}+\epsilon_{y}\mathcal{H}_{y}+\epsilon_{z}\beta_{z}}.
%\end{equation} 
%Note that $\mathcal{H}_{x}$ changes only inside a horizontal dipole since only then $R_{51}$ and $R_{52}$ of the transfer matrix are non-zero. 

%\section{Implications}

%++++++++++++++++++++++++++++++++++++++++
% References section will be created automatically 
% with inclusion of "thebibliography" environment
% as it shown below. See text starting with line
% \begin{thebibliography}{99}
% Note: with this approach it is YOUR responsibility to put them in order
% of appearance.

%

\end{document}